\documentclass[sigconf,nonacm]{acmart}

\newcommand{\herboldCount}{212}

\newcommand{\herboldLabelBugfixCount}{49}
\newcommand{\herboldLabelRefactoringCount}{1}
\newcommand{\herboldLabelDocumentationCount}{32}
\newcommand{\herboldLabelTestCount}{79}
\newcommand{\herboldLabelWhitespaceCount}{8}

\newcommand{\herboldLabelUnclearCount}{42}
\newcommand{\herboldQuestion}{Does the marked line (>>>) contribute to the bug fix, or is it something else?}

\newcommand{\hartelCount}{135}

\newcommand{\hartelLabelYesCount}{49}
\newcommand{\hartelLabelNoCount}{42}
\newcommand{\hartelLabelUnclearCount}{44}
\newcommand{\hartelQuestion}{Does this PR review comment discuss a potential security defect?}

\newcommand{\munaiahCount}{172}

\newcommand{\munaiahLabelProjectCount}{91}
\newcommand{\munaiahLabelNotprojectCount}{81}

\newcommand{\munaiahQuestion}{Is this repository an engineered software project with general-purpose utility to users other than the developers themselves?}

\newcommand{\levinCount}{129}

\newcommand{\levinLabelCorrectiveCount}{65}
\newcommand{\levinLabelAdaptiveCount}{22}
\newcommand{\levinLabelPerfectiveCount}{42}

\newcommand{\levinQuestion}{What is the maintenance intent of this commit?}

\newcommand{\totalSkipped}{108}
\newcommand{\totalRepos}{312}
\newcommand{\totalReposSkipped}{9}

\newcommand{\simpleLlmSonnetMemorizationTimeMean}{3.7s}

\newcommand{\simpleLlmSonnetMemorizationTokensOutMean}{138}

\newcommand{\simpleLlmSonnetZeroCotTimeMean}{1.5s}

\newcommand{\simpleLlmSonnetZeroCotTokensOutMean}{4}

\newcommand{\simpleLlmSonnetCostMean}{\$0.026}

\newcommand{\simpleLlmSonnetTimeMean}{5.1s}

\newcommand{\simpleLlmSonnetTokensOutMean}{189}

\newcommand{\simpleLlmLlamaCostMean}{\$0.0009}

\newcommand{\simpleLlmLlamaTimeMean}{1.9s}

\newcommand{\simpleLlmLlamaTokensOutMean}{101}

\newcommand{\simpleLlmMistralCostMean}{\$0.0044}

\newcommand{\simpleLlmMistralTimeMean}{6.0s}

\newcommand{\simpleLlmMistralTokensOutMean}{217}

\newcommand{\agentSonnetNativeCostMean}{\$0.085}

\newcommand{\agentSonnetNativeCacheSavingsPercent}{35.1\%}
\newcommand{\agentSonnetNativeTimeMean}{29.3s}
\newcommand{\agentSonnetNativeTokensInMean}{14.2K}
\newcommand{\agentSonnetNativeTokensOutMean}{1.1K}
\newcommand{\agentSonnetNativeCacheReadMean}{18.5K}
\newcommand{\agentSonnetNativeCacheWriteMean}{5.3K}
\newcommand{\agentSonnetNativeStepsMean}{10.4}

\newcommand{\agentSonnetStopseqCostMean}{\$0.067}

\newcommand{\agentSonnetStopseqCacheSavingsPercent}{40.4\%}
\newcommand{\agentSonnetStopseqTimeMean}{25.4s}
\newcommand{\agentSonnetStopseqTokensInMean}{10.7K}
\newcommand{\agentSonnetStopseqTokensOutMean}{720}
\newcommand{\agentSonnetStopseqCacheReadMean}{18.2K}
\newcommand{\agentSonnetStopseqCacheWriteMean}{4.9K}
\newcommand{\agentSonnetStopseqStepsMean}{10.5}

\newcommand{\agentMistralCostMean}{\$0.0051}

\newcommand{\agentMistralTimeMean}{21.3s}
\newcommand{\agentMistralTokensInMean}{8.5K}
\newcommand{\agentMistralTokensOutMean}{607}

\newcommand{\agentMistralStepsMean}{5.7}
\newcommand{\costRatioSonnetStopseq}{2.5}
\newcommand{\costRatioSonnetNative}{3.2}
\newcommand{\costRatioMistral}{1.2}

\newcommand{\totalSamples}{4943}
\newcommand{\totalAttempts}{5184}
\newcommand{\totalEvaluated}{648}
\newcommand{\totalSystematic}{212}
\newcommand{\totalGenuineErrors}{29}

\newcommand{\simpleLlmSonnetMemorizationNContextOverflow}{0}

\newcommand{\simpleLlmSonnetZeroCotNContextOverflow}{6}

\newcommand{\simpleLlmSonnetNContextOverflow}{7}

\newcommand{\simpleLlmLlamaNContextOverflow}{9}

\newcommand{\simpleLlmMistralNContextOverflow}{3}

\newcommand{\pBetterSonnetStopseqHartel}{67\%}

\newcommand{\pWorseSonnetStopseqHartel}{0\%}

\newcommand{\pWorseSonnetNativeHartel}{2\%}

\newcommand{\pWorseMistralHartel}{4\%}

\newcommand{\pEqualSonnetStopseqHerbold}{83\%}

\newcommand{\pEqualSonnetNativeHerbold}{86\%}

\newcommand{\pWorseMistralHerbold}{51\%}

\newcommand{\pEqualSonnetStopseqLevin}{84\%}

\newcommand{\pEqualSonnetNativeLevin}{90\%}

\newcommand{\pEqualMistralLevin}{91\%}

\newcommand{\pEqualSonnetStopseqMunaiah}{90\%}

\newcommand{\pBetterSonnetNativeMunaiah}{44\%}

\newcommand{\pBetterMistralMunaiah}{78\%}

\newcommand{\simpleLlmSonnetUnclearRate}{1\%}
\newcommand{\simpleLlmLlamaUnclearRate}{9\%}
\newcommand{\simpleLlmMistralUnclearRate}{2\%}

\newcommand{\agentSonnetNativeToolCallsPerSample}{14.1}
\newcommand{\agentSonnetNativeToolErrorRate}{1.7\%}

\newcommand{\agentSonnetStopseqToolCallsPerSample}{13.9}
\newcommand{\agentSonnetStopseqToolErrorRate}{2.2\%}

\newcommand{\agentMistralToolCallsPerSample}{6.1}
\newcommand{\agentMistralToolErrorRate}{1.9\%}

\newcommand{\disagreementCases}{254}
\newcommand{\disagreementCasesLabeled}{100}

\newcommand{\code}[1]{\texttt{#1}}

\newcommand{\unchecked}[1]{#1}

\newcommand{\fig}[0]{Fig.}
\newcommand{\tab}[0]{Table}
\renewcommand{\sec}[0]{Sec.}

\newcommand{\negvspace}[0]{\vspace{-0.60cm}}
\newcommand{\negvspacetab}[0]{\vspace{-0.07cm}}

\begin{document}

\title{Agentic Repository Mining: A Multi-Task Evaluation}
\titlenote{Accepted at the 30th International Conference on Evaluation and Assessment in Software Engineering (EASE 2026).}

\author{Johannes H{\"{a}}rtel}
\email{j.a.hartel@vu.nl}
\orcid{0000-0002-7461-2320}
\affiliation{%
  \institution{Vrije Universiteit Amsterdam}
  \country{Netherlands}
}

\renewcommand{\shortauthors}{H{\"{a}}rtel}

\begin{abstract}
Mining software repositories often requires classifying artifacts like commits, reviews,
code lines, or entire repositories into categories. Human labeling is expensive and error-prone;
limited context frequently leads to misclassifications or uncertainty in labels.
We investigate whether LLM agents that dynamically explore repositories through standard bash commands
can match the classification quality of simple LLMs that receive pre-engineered context.
Across four tasks, eight approach configurations, and \totalSamples{} classifications,
agents achieve competitive accuracy despite retrieving their own context.
The primary advantage is robustness: agents avoid context-window overflows
and scale independently of artifact size.
A manual diagnosis of \disagreementCasesLabeled{} cases where approaches disagree with the ground truth
reveals specification ambiguities and labels produced under limited context,
suggesting that accuracy against such ground truth may underestimate approaches with broader context access.
\end{abstract}




\maketitle

\section{Introduction}

Mining software repositories often requires the classification and manual labeling of artifacts like commits, reviews,
code lines, or entire repositories into categories. 
This is the empirical foundation for studies that examine such artifacts.
Examples are studies on bug fixes~\cite{DBLP:journals/ese/HerboldTLAGCBNM22}, 
security defects~\cite{DBLP:conf/ease/Hartel25}, repositories~\cite{DBLP:journals/ese/MunaiahKCN17}, 
or maintenance effort~\cite{DBLP:conf/promise/LevinY17}.

\paragraph{ML-based solutions}

Labeling by humans can be time-consuming, expensive, and error-prone.
Directly using an ML classifier is a tempting solution but needs existing labeled data
in the first place. Hybrid approaches exist, like active learning, 
that repeatedly train a classifier and then manually label promising
candidates selected by the classifier~\cite{DBLP:conf/acns/ZhangTCSLCW22,DBLP:journals/infsof/GeFQGQ22,DBLP:conf/ease/Hartel25}.
Recently, zero-shot learning with LLMs started to offer a way to mitigate initial
training data. Like a human annotator, the LLMs can be prompted with labeling guidelines 
and the artifacts to label. However, what contextual artifacts to include in the prompt
is still a challenge to human engineering~\cite{DBLP:conf/ease/AntalBFH25,DBLP:journals/corr/abs-2505-08263,DBLP:journals/corr/abs-2503-02400,DBLP:conf/msr/TafreshipourIHA25}.

\paragraph{Agentic Repository Mining}

Agentic repository mining bypasses the need for a human to pre-engineer context.
Given only a starting point like a commit SHA, the agent autonomously explores the repository
through tool calls until it gathers sufficient evidence for classification~\cite{DBLP:journals/corr/abs-2501-18160,DBLP:conf/ijcai/GuoCWCPCW024,DBLP:conf/ease/AbeduAS24}.
This removes per-task context engineering but raises the question of whether self-retrieved context
matches human-curated input.
While recent work studies agents as subjects in repositories~\cite{MSR_Robbes26,MSR_Mohsenimofidi25,MSR_TasnimDasRoy26,MSR_HoraRobbes26}, 
we use them to mine repositories.

\paragraph{Research Questions}

We evaluate such an option by putting agents into a direct comparison to a simple LLM counterpart without tools
but pre-engineered context. We target four classification tasks from related work
with different repository artifacts (lines, comments, repositories, commits).
All tasks have manually labeled ground truth; agents and LLMs get the same labeling guidelines
as in the original studies. We ask three research questions:

\begin{itemize}
    \item \textbf{RQ1:} How does agent and simple LLM accuracy compare to a human ground truth?
    \item \textbf{RQ2:} How do resource usage (tokens, time, costs) and failure modes compare between agents and simple LLMs?
    \item \textbf{RQ3:} Why do agents and simple LLMs disagree with the ground truth?
\end{itemize}

\paragraph{Tasks}
\emph{Herbold et al.}~\cite{DBLP:journals/ese/HerboldTLAGCBNM22} label whether a changed line in a bug-fixing 
commit contributes to the fix or not; 
\emph{Härtel}~\cite{DBLP:conf/ease/Hartel25} determines whether a PR review comment discusses a potential security defect;
\emph{Munaiah et al.}~\cite{DBLP:journals/ese/MunaiahKCN17} distinguish engineered software projects from homework, toy projects, or personal experiments; and
\emph{Levin et al.}~\cite{DBLP:conf/promise/LevinY17} classify commits as corrective, adaptive, or perfective.

\paragraph{Short Answers}

The major advantage of agents we find is their \textit{robustness to context size}.
Agents selectively read and thereby keep context, tokens, and cost largely independent of what artifacts are maximally available.
Simple LLMs run into context overflow errors when artifacts get large.
Agents are pricier by factors of \costRatioMistral{} to \costRatioSonnetNative{},
but pay off when artifacts exceed a certain size.
Accuracy does not show a clear winner, but agents retrieve their own context
rather than receiving it pre-engineered, making comparable accuracy a stronger result.
We diagnose \disagreementCasesLabeled{} cases where approaches disagree with the ground truth,
noticing labeling guidelines need updates. 

\paragraph{Contributions}

\begin{itemize}
    \item We introduce a simple \emph{agentic repository mining} framework to classify artifacts 
    via bash commands in isolated containers with access to the repository.

    \item We compare accuracy, resource usage, and failure modes of agents with their simple counterparts 
    across four tasks.

    \item We diagnose disagreements with the ground truth to understand where approaches fail 
    and where labeling guidelines might need to be updated.

\end{itemize}

\paragraph{Data Availability}

We provide code, prompts, and results under:

\begin{itemize}
    \item \textbf{Figshare:} \url{https://doi.org/10.6084/m9.figshare.31136470.v1}
    \item \textbf{GitHub:} \url{https://github.com/johanneshaertel/EASE_2026_agentic_repository_mining}
\end{itemize}

\paragraph{Road-Map.}
\sec~\ref{sec:background} introduces agents;
\sec~\ref{sec:approach} describes the compared approaches;
\sec~\ref{sec:tasks} presents the tasks;
\sec~\ref{sec:methodology} details our methodology;
\sec~\ref{sec:results} reports results;
\sec~\ref{sec:threats_limitations} outlines threats and limitations;
\sec~\ref{sec:related} discusses related work;
and \sec~\ref{sec:conclusion} concludes.

\section{Background}
\label{sec:background}

Context matters for classification.
In a recent postmortem, Anthropic reports expanding repository access for their code-review agents
after finding that this context was necessary to detect a critical bug~\cite{AnthropicPostmortem2026}.
Choosing the right context for classification has
been a long-standing problem in mining software repositories (MSR).

\begin{figure}
\centering
\includegraphics[width=1.00\columnwidth]{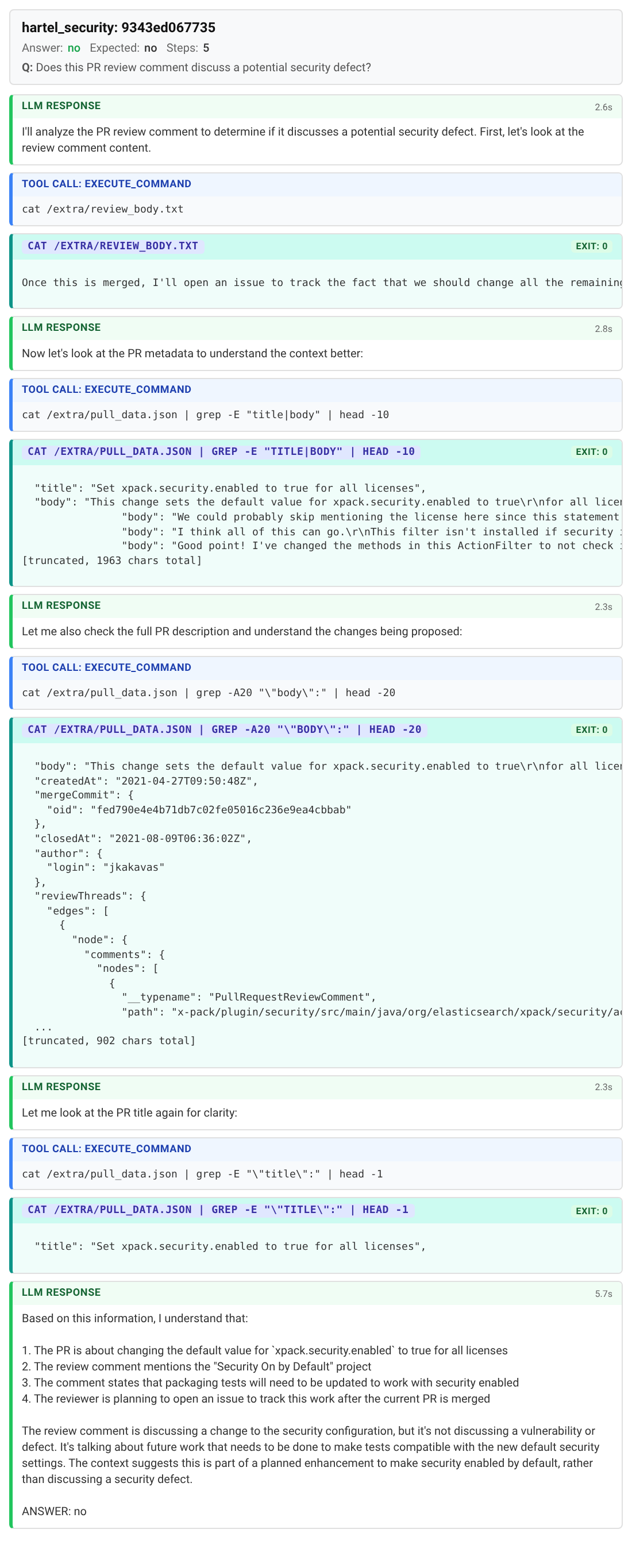}
\vspace{-3em}
\caption{Agent trajectory: Green is LLM reasoning, blue is tool calls, and teal is tool output.
The starting prompt is excluded to fit the page (see online material).}
\label{fig:trace}
\end{figure}

\subsection{Human-Engineered Context}
Human-engineered context in MSR even existed before the era of LLMs.
Established studies examine what input features machine learning
needs to make good classifications (preventing `garbage in, garbage out').
Studies examine code attributes for defect classification~\cite{DBLP:journals/tse/MenziesGF07,DBLP:conf/promise/0001PZ07,DBLP:conf/icse/ZimmermannN08},
aggregation~\cite{DBLP:journals/tse/ZhangHMZ17}, or granularity~\cite{DBLP:conf/icse/RahmanD11}.

With the rise of LLMs, context is now engineered as a prompt and passed to an LLM, but
the same core challenge applies~\cite{DBLP:journals/corr/abs-2505-08263,DBLP:journals/corr/abs-2503-02400,DBLP:conf/msr/TafreshipourIHA25}.
What context needs to be added to the prompt to get better classifications?
A recent example by Antal et al. is adding metadata on security defects to the prompt to improve performance~\cite{DBLP:conf/ease/AntalBFH25}.

\subsection{Human Context Retrieval}

If humans are to classify an artifact, they may follow a special strategy, not looking
at fixed numbers of lines or code attributes of the artifact, like classical ML approaches would do~\cite{DBLP:conf/icse/SpinellisG18}.

For example, if a classification depends on precise knowledge of a class at a certain revision, humans may
clone the repository, check out the relevant revision, and search for the class in the working copy.
This might reveal another class with more critical information, a class that might be difficult to anticipate beforehand.
In a nutshell, such a sequence of bash commands, e.g., \code{git clone}, \code{git checkout}, and \code{rg ClassName},
has two important properties: 

\begin{itemize}
    \item First, it is \textbf{autoregressive}. Classification is done in multiple steps, and each step depends on previous results.
    There is no fixed vector of information but a growing sequence. A threshold might constrain exploration depth.

    \item Second, it uses \textbf{existing development tools}, in our example, bash access to explore the repository.
\end{itemize}

\subsection{Agentic Context Retrieval}

Agentic repository mining builds on the same strategy.
We show an example from a run of our study in \fig~\ref{fig:trace} where an agent extracts the necessary
context by a cascade of bash commands to dynamically explore what is needed for the classification.
The agent iteratively inspects files, searches code, and examines the history until it gathers sufficient evidence
and concludes in the final step.
Standard bash commands are the universal tool used; no specialized API or retrieval system is needed.
Systematic evaluations of agents used for MSR are still in early stages~\cite{DBLP:journals/corr/abs-2501-18160,DBLP:conf/ijcai/GuoCWCPCW024,DBLP:conf/ease/AbeduAS24}.

\fig~\ref{fig:trace} shows details like how selectively the agent explores the available content, only recovering
the particular lines of the files needed to make the classification, not reading the entire files or repository.
This is done by piping the results of \code{cat} into \code{grep} and \code{head}.
The agent is aware of the schema of GitHub's PR JSON data, which is available as a file dump, as it knows there is 
just one title (see \code{head -1}).
We provide such traces for all our \totalSamples{} experiments.

\subsection{Agents in a Nutshell}
Agentic approaches build upon LLMs using the same autoregressive text generation.
The LLM predicts the next token given all previous tokens, iterating until a stop sequence is met.
A simple invocation of an LLM typically generates text until it produces an answer.

Agents differ from standard LLM calls by interleaving generation with tool execution.
An agent specializes the loop by generating until the generated text conforms to a valid tool invocation, like \code{... <bash> ls </bash>}.
When the model outputs such a tool call, regular LLM-based generation pauses; in our example, by the stop sequence \code{</bash>}.
The tool \code{ls} is executed, the tool output (listing of files and directories) 
is appended to the growing string, and the LLM resumes generation conditioned on the entire text so far,
including the new tool output. This is done until the next tool call is generated or a final answer is produced.
This interleaving of LLM generation and tool execution enables the dynamic context recovery shown in \fig~\ref{fig:trace}.
We refer to the online appendix for our simplistic implementation of an agent that we use in our experiments.

The simplicity of Bash is not a limitation, being the universal tool call that orchestrates other tools.
Our experiments focus on bash calls to \code{git} but extensions to what can be called are straightforward.







\section{Approaches}
\label{sec:approach}

We compare simple LLM baselines with fixed pre-engineered context 
against an agentic harness on top of the same LLMs
that enables dynamic context exploration via tool execution logic.
For simple LLM invocations, we evaluate
chain-of-thought, no chain-of-thought, and a pure memorization baseline without any context.
For agents, we evaluate two tool-calling variants. 

\subsection{Base Prompt}
All compared approaches get labeling guidelines close to the human annotators in the original studies.
This includes the classification question and the list of categories that are possible answers. 
We allow models to also answer with \textit{unclear} as an additional category rather than guessing.
What follows is specific to the approaches.

\subsection{Simple LLMs}

Simple LLM approaches are executed in a single turn until a final answer is produced.
Therefore, additional context must be appended after the base prompt before passing it to the LLM. 
This needs to be engineered by us adding resources like commit diffs, file contents, or PR
descriptions specific to each task.
We keep this close to the original studies with partial deviations that
we discuss in Sec.~\ref{sec:tasks}.

\subsubsection{Memorization}

We evaluate classification without context, where we only append the minimal identification of central artifacts
to see if the LLM can recall the correct label anyhow.
Such an effect would be a strong indication that either the context or the final labels are memorized 
because they are in the training data. We avoid the category `unclear' for this variant.

\subsubsection{Chain-of-Thought}

We evaluate chain-of-thought prompting in two modes: with and without reasoning.
The standard variant is instructed to produce a reasoning trace followed by the answer.
The no-chain-of-thought variant (no-CoT) is explicitly instructed to produce
only the final answer without any reasoning trace. We enforce this by a strong prompt
that leads to a direct answer.

\subsection{Agents}

Our agentic approaches dynamically append context themselves in the agentic loop, having access to a shell executed in a 
sandboxed container where the repository is cloned.
There is no need for task-specific engineering of the context.
The sandboxed Docker environment ensures reproducibility and safety: each run starts from 
a clean repository state, and the agent cannot affect the host system.
The container does not have internet access.

The container also includes additional information as files that are not directly available via \code{git}
to align with what is given to the simple LLMs.
This is data that would typically be retrieved over internet APIs that we did not allow the agents to have access to.
Compared to the simple LLM approaches, such a setup keeps the starting prompt much smaller but
does not limit the agents' information to a subset of what is available to the simple LLMs.
We limit agents to 50 exploration steps to bound cost and runtime
and explicitly record if the agent runs into the step limit.

\subsubsection{Tool Calling}

We examine two variations in tool calling:

\begin{itemize}
\item \textbf{Native}: Modern LLM APIs provide native tool calling where the model can output structured tool invocations.
Such APIs handle parsing specific to the tools and native to the model. 
It is a vendor-specific, standardized, but less transparent way of tool calling.
We use the Converse API from Amazon Bedrock for this, 
limited to one native tool call per step (boto3, v1.40.15, \url{https://pypi.org/project/boto3/}).
This call might contain multiple chained and piped bash commands.

\item \textbf{Stop-Sequence}: We also implement the 
stop-sequence solution we have described in \sec~\ref{sec:background}. The model writes 
tool calls as XML (\texttt{<bash> command </bash>}), the closing tag stops generation, we extract the command using
basic string operations, and interpret the content as a bash call. 
This approach is API-agnostic, transparent, and simple, since the full interaction is visible in plain text.
\end{itemize}

\subsubsection{Prompt Caching}

In contrast to simple LLM invocations, agentic interactions involve many turns, repeatedly resending the system 
prompt, and a growing conversation history.
Transformer-based LLMs, as exposed through APIs such as Bedrock, are stateless across calls 
and predict the next token solely from the provided context.
There is no persistent session state between invocations.
As a consequence, token usage and associated costs grow with conversation length, 
which is a major factor when evaluating agents.

Prompt caching mechanisms can reduce these costs by marking prompt prefixes that persist across separate calls.
When such prefixes are reused, they are billed at reduced rates, while model behavior remains unchanged.
Among the selected models, prompt caching is supported only by Claude~3.7~Sonnet via Bedrock.
For the stop-sequence agent, we implement a manual caching scheme along a regular grid.
For native tool calling, we rely on the Bedrock prompt caching over the Converse interface of boto3.

\subsection{Models}

We evaluate three models via Amazon Bedrock. All models are examined in simple LLM mode with chain-of-thought;
the memorization and no-CoT variants are run with Claude 3.7 Sonnet only.
Only Claude 3.7 Sonnet and Mistral Large 3 are used for agents. Mistral is limited to native tool calling.

\begin{itemize}
    \item \textbf{Claude 3.7 Sonnet}: 
    The vendor reports a context window of about 200K tokens. One million input tokens cost
    3.0 USD, output tokens 15.0 USD, cached reads 0.3 USD, and cached writes 3.75 USD.
    In January 2026, only Sonnet 3.7 reliably used the
    stop-sequence tool calling in dry runs, so the other models have been excluded from this approach.

    \item \textbf{Mistral Large 3}: One million input tokens cost 0.5 USD, and output tokens cost 1.5 USD. 
    The context size is 256K tokens. Prompt caching is not supported.

    \item \textbf{Llama 3.3 70B}: Llama did not show reliable tool-calling 
    in our dry runs, so we excluded it from the agents. The context size is 128K tokens.
    One million input tokens cost 0.15 USD, output tokens 0.6 USD. Prompt caching is not supported.
\end{itemize}

Llama is selected as a popular open model, Mistral as a model that balances performance and costs,
and Claude 3.7 Sonnet as the most expensive model with strong tool-calling capabilities.

\section{Tasks}
\label{sec:tasks}

We evaluate the approaches on four MSR classification tasks from prior work, each with a manually labeled ground truth.
The tasks vary in granularity, from entire repositories to individual lines.
\tab~\ref{tab:tasks} summarizes the tasks and the context provided to each approach.
We refer to the datasets by the first author of the papers.

\begin{table}[t]
\centering
\caption{Classification tasks: Samples \textbf{N};
context shows what each approach receives; simple LLMs only\textsuperscript{S}, agent only\textsuperscript{A}.}
\label{tab:tasks}
\negvspacetab
\small
\begin{tabular}{@{}llrp{4.3cm}@{}}
\toprule
\textbf{Task} & \textbf{Unit} & \textbf{N} & \textbf{Context} \\
\midrule
Munaiah & Repo & \munaiahCount{} & Dir.\ listing\textsuperscript{S}, README\textsuperscript{S}; full repo\textsuperscript{A} \\
\multicolumn{4}{@{}p{0.97\columnwidth}@{}}{\scriptsize \textbf{Q:} \munaiahQuestion\ --- project, notproject, unclear} \\[6pt]
Herbold & Line & \herboldCount{} & Diff hunk w/ marked line; full repo\textsuperscript{A} \\
\multicolumn{4}{@{}p{0.97\columnwidth}@{}}{\scriptsize \textbf{Q:} \herboldQuestion\ --- bugfix, test, doc., refactoring, whitespace, unrelated, unclear} \\[6pt]
H\"artel & Review & \hartelCount{} & Comment + PR metadata; full repo\textsuperscript{A} \\
\multicolumn{4}{@{}p{0.97\columnwidth}@{}}{\scriptsize \textbf{Q:} \hartelQuestion\ --- yes, no, unclear} \\[6pt]
Levin & Commit & \levinCount{} & Commit message; full repo\textsuperscript{A} \\
\multicolumn{4}{@{}p{0.97\columnwidth}@{}}{\scriptsize \textbf{Q:} \levinQuestion\ --- corrective, adaptive, perfective, unclear} \\
\bottomrule
\end{tabular}
\end{table}

\subsection{Herbold et al.: Tangled Commits}

Bug-fixing commits frequently contain unrelated changes, like 
refactoring, documentation updates, or whitespace cleanup.
These \emph{tangled commits} introduce noise in defect prediction 
studies that assume all lines in a bug-fix commit address the bug~\cite{DBLP:journals/ese/HerboldTLAGCBNM22}.

Herbold et al. created a dataset with line-level labels for changes in bug-fixing commits from 28 Java projects.
Multiple annotators independently classified each changed line.
We use \herboldCount{} sampled lines from this dataset with \herboldLabelBugfixCount{} bugfix,
\herboldLabelTestCount{} test, \herboldLabelDocumentationCount{} documentation,
\herboldLabelWhitespaceCount{} whitespace, \herboldLabelRefactoringCount{} refactoring, and \herboldLabelUnclearCount{} unclear.
We interpret lines with no consensus in the original study as unclear lines.

Simple LLMs and agents receive the diff hunk with the target line marked.
The agent also has access to the full repository.

\subsection{H{\"{a}}rtel: Security Reviews}

Code review is an opportunity to catch security defects before they reach production.
Previous work of ours identifies which reviews discuss potential security defects to help prioritize fixes and understand 
how security issues surface during development~\cite{DBLP:conf/ease/Hartel25}.

Previous work used active learning with a fine-tuned language model to efficiently mine 
and classify four million GitHub PR reviews.
For this work, we use a balanced sample of \hartelCount{} reviews from the original data, with
\hartelLabelYesCount{} discussing a potential security defect,
\hartelLabelNoCount{} not discussing one, and \hartelLabelUnclearCount{} unclear cases.

The simple LLMs and agents receive the review comment text and PR metadata accessible via the GitHub API,
including the full discussion thread.
The agent also has access to the full repository.

Our previous study running on four million reviews did not pass the full metadata 
to the fine-tuned LLM for scalability reasons, which renders the available context different.
Differences in context may affect classification, which we discuss in the results (\sec~\ref{sec:results}).

\subsection{Levin et al.: Maintenance Activities}

Software maintenance activities can be classified into three categories~\cite{DBLP:conf/icsm/MockusV00}: 
corrective, adaptive, and perfective.
Automatically classifying commits by such type supports project assessment.

Levin et al. manually labeled 1151 commits from 11 popular open-source projects~\cite{DBLP:conf/promise/LevinY17}
to eventually train classifiers.
We use \levinCount{} of the provided manually labeled commits with \levinLabelCorrectiveCount{} corrective, 
\levinLabelAdaptiveCount{} adaptive, and \levinLabelPerfectiveCount{} perfective.

The simple LLM receives only the commit message.
The agent has access to the full repository.

\subsection{Munaiah et al.: Repository Classification}

GitHub hosts many repositories that are homework, assignments, backups, empty projects, or personal experiments.
This is a problem for MSR studies that aim to generalize to \emph{engineered software projects} rather than toy or personal projects~\cite{DBLP:journals/ese/MunaiahKCN17}.

Munaiah et al. trained classifiers on extractable repository features such as architecture, community activity, and continuous integration.
For validation, two authors independently labeled 200 repositories; only consensus cases are included.
We use the \munaiahCount{} repositories still accessible (\munaiahLabelProjectCount{} project, \munaiahLabelNotprojectCount{} not-project).

The simple LLM receives the directory listing and README file appended to the prompt.
The agent has access to the full repository.
\section{Methodology}
\label{sec:methodology}

\smallskip

\subsection{Sample Selection}
\label{sec:sample-selection}

For Munaiah, we use \munaiahCount{} accessible repositories of 200 originally labeled.
For Herbold, we randomly sample 300 lines with at least three annotators; 
lines without consensus ($\geq$75\% agreement) are labeled unclear; \herboldCount{} remain after exclusions.
For H{\"{a}}rtel, we draw a stratified sample of 50 per class; \hartelCount{} remain.
For Levin, we randomly sample 200 commits; \levinCount{} remain.
We exclude \totalSkipped{} observations across \totalReposSkipped{} of \totalRepos{} repositories 
due to excessive repository size (above 1000 MB).
Herbold misses cases where we did not manage to verify the correct line and label association from the original dataset.
Exclusions and filters apply to all approaches to ensure a fair comparison.

\subsection{Prompts}
\label{sec:prompt-design}

Agentic and non-agentic approaches get closely matched prompts to ensure a fair comparison.
We describe the details of prompt construction in \sec~\ref{sec:approach} and \tab~\ref{tab:tasks}.
Full prompts can also be found in the replication package,
where they are stored with the trajectories for each experiment.
For validation, we sampled 20 experiments to confirm correct prompt construction.


\subsection{Execution}
\label{sec:execution}

We run experiments across four tasks and eight approaches
on \totalEvaluated{} samples per approach (after size exclusions, see \sec~\ref{sec:sample-selection}).
This yields \totalAttempts{} attempts, of which \totalSystematic{} are systematic
(memorization does not apply to Herbold lines) and \totalGenuineErrors{} are
genuine errors (mostly context overflow on simple LLMs; zero for agents),
leaving \totalSamples{} valid classifications.
Models are accessed via Amazon Bedrock to avoid infrastructure differences.
We set the temperature to 1.0 and the maximum output tokens to 4096 for all approaches except the no-CoT variant (16 tokens).
Top-p and other sampling parameters use Bedrock defaults.
The experiments were executed in April 2026.


\subsection{Measurements}
\label{sec:measurements}

We measure classification accuracy, resource usage (tokens, time, cost),
exploration steps, and tool-use behavior to answer our research questions RQ1 and RQ2.
Errors of the approaches are recorded and classified.
Cost is estimated from the tokens using the prices of AWS US East (N. Virginia) as of January 2026.
Such costs are a reasonable approximation of the resource usage behind the APIs.


\subsection{Disagreement Analysis}
\label{sec:disagreement-procedure}

To understand why approaches disagree with the ground truth and answer RQ3, we
identify all cases where at least two non-experimental approaches disagree
with the ground-truth, yielding \disagreementCases{} such cases
across the tasks. We manually inspect a stratified sample of \disagreementCasesLabeled{} cases and assign 
one of four diagnoses from \tab~\ref{tab:disagreement-taxonomy}.

\begin{table}[t]
    \centering
    \caption{Taxonomy for diagnosing disagreements}
    \label{tab:disagreement-taxonomy}
    \negvspacetab
    \small
    \begin{tabular}{@{}lp{6cm}@{}}
    \toprule
    Diagnosis & Definition \\
    \midrule
    Update label  & Evidence supports a different label than the original ground truth; the ground truth should be revised. \\
    Keep label    & The original ground truth holds; the disagreeing approaches misclassify. \\
    Specification & The task definition is ambiguous; both (multiple) labels are defensible interpretations. \\
    Unresolvable  & Insufficient information to determine the correct label. \\
    \bottomrule
    \end{tabular}
\end{table}

\subsection{Hierarchical Accuracy Model}

We use a hierarchical accuracy model to estimate approach
accuracy with credible intervals. The model accounts for per-sample
difficulty, yielding uncertainty that reflects the paired design
where all approaches classify the same samples. For a single task, the model
is specified as follows:
\begin{align}
  y_{s,a} &\sim \mathrm{Bernoulli}(\mathrm{logit}^{-1}(\alpha_a + \theta_s)) \\
  \theta_s &\sim \mathrm{Normal}(0, \sigma_\theta) \\
  \alpha_a &\sim \mathrm{Normal}(0, 2) \\
  \sigma_\theta &\sim \mathrm{HalfNormal}(2)
\end{align}
where $y_{s,a} \in \{0,1\}$ indicates whether approach~$a$ correctly
classified sample~$s$, $\alpha_a$ captures approach ability, and
$\theta_s$ captures sample difficulty. Errors and timeouts can be excluded
from the likelihood by setting those values to NaN.
We fit one model per task using Stan~\cite{stan}.
Approach accuracy is derived as the posterior predictive mean
$\mathrm{acc}_a = S^{-1} \sum_s \mathrm{logit}^{-1}(\alpha_a + \theta_s)$.
Pairwise accuracy differences $\mathrm{acc}_a - \mathrm{acc}_b$ are computed
directly from posterior samples, yielding credible intervals on differences.
The online material shows validation of the model 
by fitting it to simulated data (see~\cite{DBLP:journals/ese/HartelL23,MorrisWC19}). 


\section{Results}
\label{sec:results}

\smallskip

\subsection{Resource Usage}
\label{sec:resource-usage}

\smallskip

\begin{figure}
    \centering
    \includegraphics[width=\columnwidth]{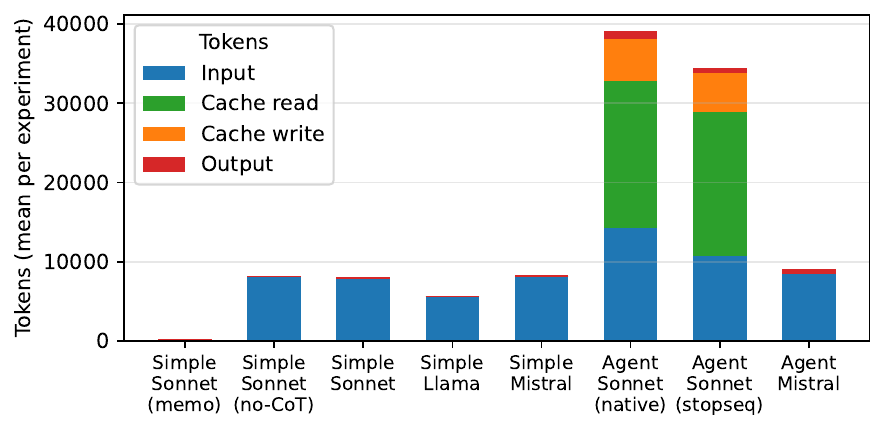}
    \negvspace
    \caption{Mean token usage by approach over all experiments.}
    \label{fig:token-barplot}
\end{figure}

\paragraph{Tokens.}
\fig~\ref{fig:token-barplot} shows the token composition per experiment.
Simple LLM approaches, except for memorization, use around \unchecked{5--8K} input tokens independent of the underlying model. 
Numbers are different because context overflow errors are excluded from the mean
(and from cost aggregates), removing the highest token counts from the aggregation.

The occurrence of context overflow errors depends on the model's context limit and the specific prompt size.
Llama has the smallest context size limit of 128K tokens, Sonnet lies around 200K tokens,
and Mistral has the largest with 256K tokens.
Simple Llama runs into \simpleLlmLlamaNContextOverflow{} context-overflow errors,
simple Sonnet into \simpleLlmSonnetNContextOverflow{},
simple Sonnet (no-CoT) into \simpleLlmSonnetZeroCotNContextOverflow{},
and simple Mistral into \simpleLlmMistralNContextOverflow{}.
Simple Sonnet (memorization) runs into \simpleLlmSonnetMemorizationNContextOverflow{} errors
because the approach does not get context that could exceed the limit.

Simple LLMs only output a reasoning trace and answer, no exploration steps or tool calls,
so output tokens are low.
Llama outputs on average \simpleLlmLlamaTokensOutMean{} tokens,
Mistral \simpleLlmMistralTokensOutMean{},
simple Sonnet \simpleLlmSonnetTokensOutMean{},
simple Sonnet (memorization) \simpleLlmSonnetMemorizationTokensOutMean{},
and simple Sonnet (no-CoT) \simpleLlmSonnetZeroCotTokensOutMean{} tokens.

Agentic approaches show an entirely different picture because
tokens accumulate over multiple steps, and thereby numbers are much higher.
Overall, Agent Mistral uses \agentMistralTokensInMean{} fresh input tokens on average
with \agentMistralTokensOutMean{} output tokens for reasoning traces and tool calls.
Agent Sonnet (stopseq) gets \agentSonnetStopseqTokensInMean{} fresh input tokens on average
plus \agentSonnetStopseqCacheReadMean{} cache-read tokens, \agentSonnetStopseqCacheWriteMean{} cache-write tokens,
with \agentSonnetStopseqTokensOutMean{} output tokens for reasoning traces and tool calls.
Agent Sonnet (native) uses the most tokens overall with \agentSonnetNativeTokensInMean{} fresh input tokens on average
plus \agentSonnetNativeCacheReadMean{} cache-read tokens, \agentSonnetNativeCacheWriteMean{} cache-write tokens,
and \agentSonnetNativeTokensOutMean{} output tokens for reasoning traces and tool calls.

We see a significant amount of cached read tokens for both Sonnet approaches.
Mistral does not support caching.
Agents run into no context-window overflows,
which is a major advantage in the robustness of agentic approaches.
The simplistic stop-sequence approach appears to be more efficient in token use than
native tool calling. This is likely because the stop-sequence approach
cuts generation after the tool call is determined, which
accelerates the feedback loop between tool output and next LLM generation.

\paragraph{Cost.}
We next analyze the cost per experiment because it correlates with the effective resource usage of each approach.
Due to the different pricing schemes of the models, cost is not a one-to-one mapping from tokens.
Provider costs serve as a practical proxy for resource use rather than a scientific end in themselves.

\begin{figure}
    \centering
    \includegraphics[width=\columnwidth]{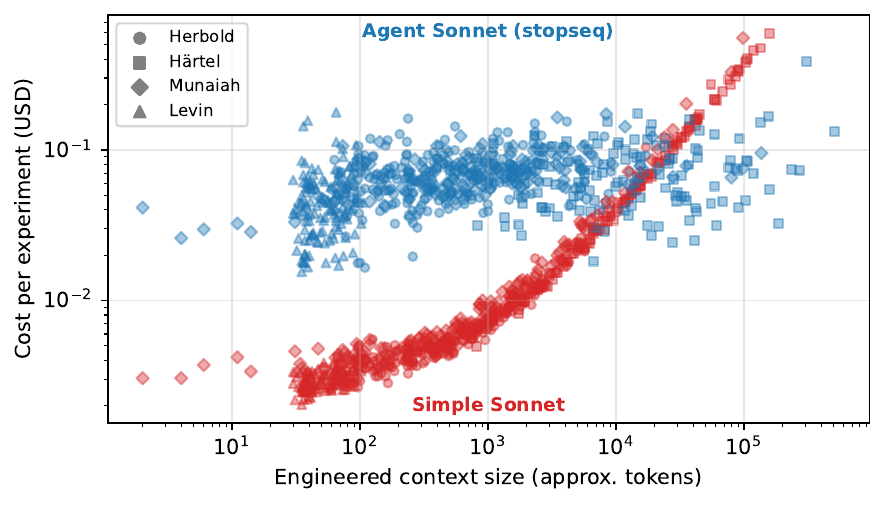}
    \negvspace
    \caption{Cost per experiment vs.\ engineered context size}
    \label{fig:context-scatter}
\end{figure}

\fig~\ref{fig:context-scatter} reveals a fundamental difference in how the two paradigms
scale. The engineered context size on the x-axis depicts the maximal input for an experiment, 
including all possible artifacts we feed into a simple approach.
Thereby, the engineered context size almost perfectly correlates with the cost for a simple approach shown on the y-axis.
Large READMEs, metadata dumps, or diffs directly cause the prompt to grow and increase the cost (red).
Agent costs, by contrast, show near-zero correlation (blue).

\begin{figure}
    \centering
    \includegraphics[width=\columnwidth]{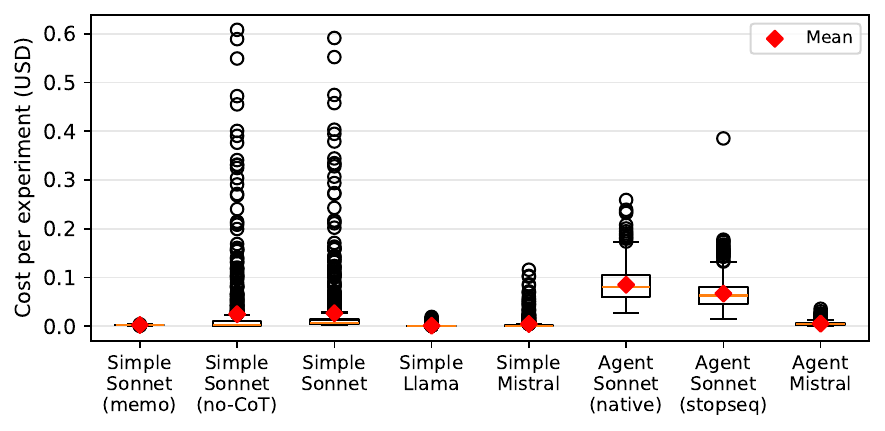}
    \negvspace
    \caption{Mean cost distribution per experiment by approach: Diamond markers (mean); boxes (median and quartiles).}
    \label{fig:cost-boxplot}
\end{figure}

\fig~\ref{fig:cost-boxplot} shows a corresponding cost distribution of all approaches.
Simple approaches are dominated by outliers of
exceptionally large inputs, but they are still the cheapest 
when averaged over all of our experiments.
Simple Llama needs \simpleLlmLlamaCostMean{}/experiment,
simple Mistral \simpleLlmMistralCostMean{}/experiment,
and simple Sonnet \simpleLlmSonnetCostMean{}/experiment.

Agents incur higher costs on average, but by small factors when compared to the corresponding simple approaches.
Agent Sonnet (native) needs an average of \agentSonnetNativeCostMean{}/experiment (factor \costRatioSonnetNative{}),
Agent Sonnet (stopseq) \agentSonnetStopseqCostMean{}/experiment (factor \costRatioSonnetStopseq{}),
and Agent Mistral \agentMistralCostMean{}/experiment (factor \costRatioMistral{}).
Especially the low increase of costs for Mistral is surprising.
Moderate cost increases can be explained by the selective exploration of agents
that keeps the overall context size manageable. 
For experiments where the engineered context exceeds approximately 30K tokens,
Agent Sonnet (stopseq) becomes more cost-effective than Simple Sonnet, as shown in \fig~\ref{fig:context-scatter}.
Hence, with a different distribution of input sizes, agents
might even be cheaper on average than simple approaches.

Caching helps reduce input costs; cached reads are billed at 10\% of the regular input price. 
For Sonnet native, we see \agentSonnetNativeCacheSavingsPercent{} cost savings through caching,
and for Sonnet stopseq, \agentSonnetStopseqCacheSavingsPercent{} savings.
Our caching implementation has not been optimized, and we expect bigger savings with more advanced caching strategies.

\paragraph{Time.}
Simple LLM approaches complete fast:
simple Llama runs in \simpleLlmLlamaTimeMean{},
simple Mistral in \simpleLlmMistralTimeMean{},
simple Sonnet in \simpleLlmSonnetTimeMean{},
simple Sonnet (no-CoT) in \simpleLlmSonnetZeroCotTimeMean{},
and simple Sonnet (memorization) in \simpleLlmSonnetMemorizationTimeMean{} per experiment.
Agents naturally take longer due to sequential tool execution:
Agent Sonnet (native) takes on average \agentSonnetNativeTimeMean{},
Agent Sonnet (stopseq) \agentSonnetStopseqTimeMean{},
and Agent Mistral \agentMistralTimeMean{} per experiment.
Overall, we consider time as a secondary argument for our use case, since such
classification tasks can be run in parallel and offline.

\subsection{Tool Use}
\label{sec:tool-use}

Agents explore repositories through bash commands.
Overall, Agent Sonnet (native) shows an average of \agentSonnetNativeToolCallsPerSample{} commands 
per experiment, with an exploration depth of \agentSonnetNativeStepsMean{} steps;
Agent Sonnet (stopseq) needs \agentSonnetStopseqToolCallsPerSample{} commands on \agentSonnetStopseqStepsMean{} steps;
Agent Mistral needs \agentMistralToolCallsPerSample{} commands on \agentMistralStepsMean{} steps.
Commands higher than the step count indicate composition by piping or chaining
in the single bash invocation the agents have per step.

\begin{figure}
    \centering
    \includegraphics[width=\columnwidth]{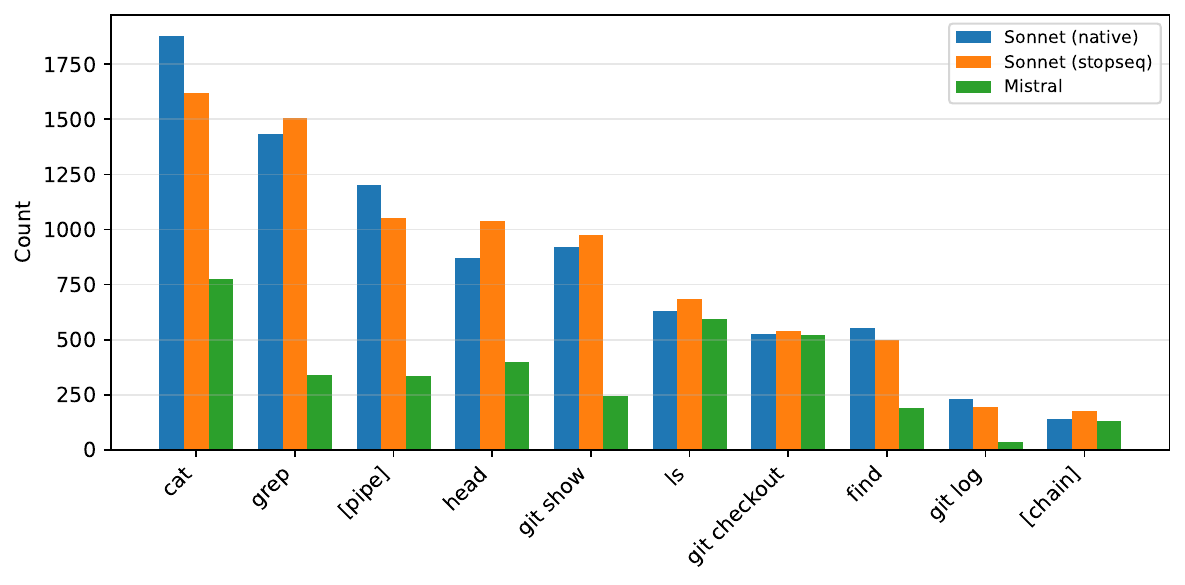}
    \negvspace
    \caption{Tool usage by agent approach (top 10 commands).}
    \label{fig:tool-usage}
\end{figure}

\fig~\ref{fig:tool-usage} shows a breakdown of tool usage for agents.
The most frequent commands are file inspection and ways to limit reads (\code{cat}, \code{head}, \code{grep}),
navigation (\code{ls}, \code{find}),
and Git operations (\code{git show}, \code{git checkout}).
Agents compose commands differently.
Sonnet pipes commands (\code{|}) preferentially, while Mistral also uses chaining a lot (\code{\&\&}, \code{;}).
We consider pipes as a more advanced way of composing commands, like using \code{grep} to filter \code{ls} output directly.
This suggests the exploration of Sonnet is more sophisticated.

\begin{table}
    \centering
    \caption{Error overview by approach}
    \negvspacetab
    \label{tab:tool-failures}
    \small
\begin{tabular}{@{}l r rr@{}}
\toprule
 & \textbf{Non-fatal} & \multicolumn{2}{c}{\textbf{Fatal}} \\
\cmidrule(lr){2-2} \cmidrule(l){3-4}
 & \shortstack{Tool\\Errors} & \shortstack{Invalid\\Category} & \shortstack{Context\\Overflow} \\
\midrule
Simple Sonnet (memo) & -- & 3 (0.5\%) & -- \\
Simple Sonnet (no-CoT) & -- & -- & 6 (0.9\%) \\
Simple Sonnet & -- & -- & 7 (1\%) \\
Simple Llama & -- & 1 (0.2\%) & 9 (1\%) \\
Simple Mistral & -- & -- & 3 (0.5\%) \\
Agent Sonnet (native) & 1.7\% & -- & -- \\
Agent Sonnet (stopseq) & 2.2\% & -- & -- \\
Agent Mistral & 1.9\% & -- & -- \\
\bottomrule
\end{tabular}
\end{table}

\tab~\ref{tab:tool-failures} summarizes errors across all approaches.
Tool errors are classified by exit codes and command outputs; fatal errors with context overflow
and invalid category invalidate the experiment and may be counted as wrong 
classifications depending on the analysis.
Tool-calling errors on the bash level are non-fatal, as the agents can recover from them in the next step.
All agents show comparable tool call error rates (\agentSonnetNativeToolErrorRate{}, \agentSonnetStopseqToolErrorRate{}, \agentMistralToolErrorRate{} for Sonnet native, stopseq, and Mistral).
Invalid category outputs are rare overall, occurring only for simple Llama and
simple Sonnet (memorization) with a total of 4 cases across all experiments.

\subsection{Exploration Depth}
\begin{figure}
    \centering
    \includegraphics[width=\columnwidth]{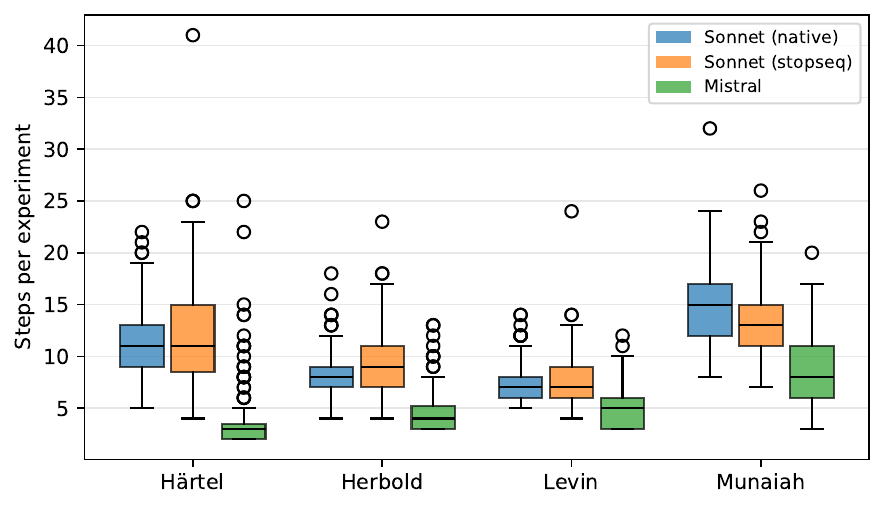}
    \negvspace
    \caption{Boxplot of steps per experiment by task and agent}
    \label{fig:steps-per-task}
\end{figure}

\fig~\ref{fig:steps-per-task} shows the distribution of steps that the agents need for each task.
We see there is variation over tasks, but overall, Sonnet approaches consistently use more steps 
than Mistral, with \agentSonnetNativeStepsMean{}, \agentSonnetStopseqStepsMean{}, and \agentMistralStepsMean{} 
steps on average for Sonnet native, stopseq, and Mistral.

Munaiah et al. (repository classification) and H\"artel (security defects in reviews) require more exploration steps
while Levin (commit classification) and Herbold (tangled commits) require fewer steps. We
assume that commits are more self-contained, while repository and review classification
require more context exploration.
No agent hits the 50-step limit.

\subsection{Classification Accuracy}
\label{sec:accuracy}

\begin{figure}
    \centering
    \includegraphics[width=\columnwidth]{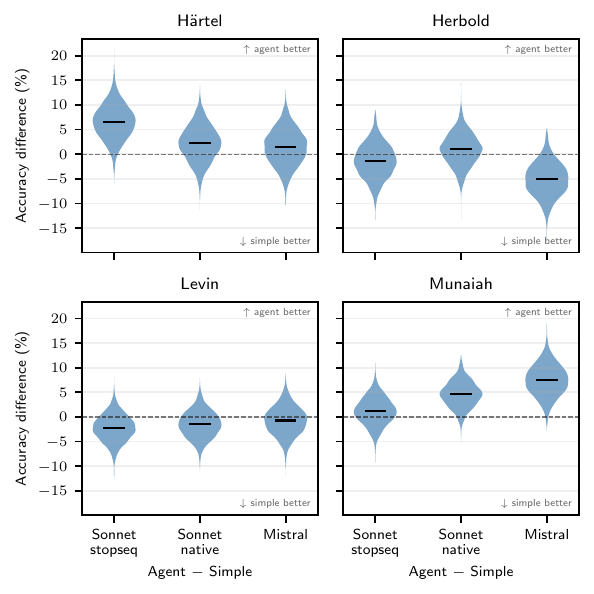}
    \negvspace
    \caption{Posterior of the accuracy difference as violin
    with fatal errors counted as incorrect classification.}
    \label{fig:accuracy-diff}
\end{figure}

We report on absolute accuracy specific to tasks in 
\tab~\ref{tab:accuracy-herbold}, \ref{tab:accuracy-hartel}, \ref{tab:accuracy-levin}, and ~\ref{tab:accuracy-munaiah}.
Statistical analysis of the accuracy difference between agents and corresponding simple approaches 
can be found in \fig~\ref{fig:accuracy-diff}.
For Agent Sonnet native and stopseq, we use Simple Sonnet as the reference. 
For Agent Mistral, we use Simple Mistral as the reference.

\begin{table}
    \centering
    \caption{Herbold (tangled commits): 95\% credible intervals;
    \textit{Acc.} excludes or includes errors;
    invalid categories (Inv.\ Cat.) and context overflow (Ctx.\ Ovfl)}
    \label{tab:accuracy-herbold}
    \negvspacetab
    \footnotesize
    \setlength{\tabcolsep}{2pt}
\begin{tabular}{lrrrr}
\toprule
\textbf{Approach} & \textbf{Acc.} & \shortstack{\textbf{Acc.}\\err=fail} & \shortstack{Inv.\\Cat.} & \shortstack{Ctx.\\Ovfl} \\
\midrule
Simple Sonnet (no-CoT) & 57.5\% {\tiny[53,62]} & $=$ & -- & -- \\
Simple Sonnet & 55.2\% {\tiny[51,60]} & $=$ & -- & -- \\
Simple Llama & 46.2\% {\tiny[42,51]} & $=$ & -- & -- \\
Simple Mistral & 54.7\% {\tiny[50,59]} & $=$ & -- & -- \\
Agent Sonnet (native) & 56.1\% {\tiny[51,61]} & $=$ & -- & -- \\
Agent Sonnet (stopseq) & 53.8\% {\tiny[49,58]} & $=$ & -- & -- \\
Agent Mistral & 49.5\% {\tiny[45,54]} & $=$ & -- & -- \\
\midrule
Random prediction\textsuperscript{$\dagger$} & 14.3\% & 14.3\% & -- & -- \\
Majority prediction\textsuperscript{$\ddagger$} & 37.3\% & 37.3\% & -- & -- \\
\bottomrule
\multicolumn{5}{@{}l}{\tiny $\dagger$\,Uniform random ($K\!=\!7$). $\ddagger$\,Always predicts the most frequent label.}
\end{tabular}
\end{table}

\subsubsection{Herbold}
\tab~\ref{tab:accuracy-herbold} shows that tangled commit line classification appears to be the hardest task for all approaches,
with accuracy values between 46.2\% and 56.1\%. No approach runs into fatal errors, so there is no
separate accuracy that needs to be reported. Testing no context with Sonnet (memorization) is excluded 
because we did not see a reasonable way to identify a specific line without
repeating the actual diff in the context.

All approaches are better than random prediction with 14.3\% accuracy or
majority prediction with 37.3\% accuracy.
Overall, Mistral and Llama approaches are slightly worse than Sonnet approaches.
We explore the corresponding pairs of agents and simple
counterparts in \fig~\ref{fig:accuracy-diff}.
Using a region of practical equivalence~\cite{kruschke2014doing,kruschke2018rejecting} of $\pm$5 accuracy points
in difference, both Sonnet agents are practically equivalent to Simple Sonnet
(\pEqualSonnetStopseqHerbold{} and \pEqualSonnetNativeHerbold{} probability).
Agent Mistral is the only case where the posterior leans toward a meaningful deficit
(\pWorseMistralHerbold{} probability of being practically worse than Simple Mistral),
though the evidence is weak.

Since no approach runs into errors on this task, agents cannot benefit from
their robustness advantage here, which leaves a pure accuracy comparison.
Being practically equal to simple approaches, despite having to find context themselves, is
the result for this task.

\begin{table}
    \centering
    \caption{H\"artel (security defects, columns as in \tab~\ref{tab:accuracy-herbold}).}
    \label{tab:accuracy-hartel}
    \footnotesize
    \negvspacetab
    \setlength{\tabcolsep}{2pt}
\begin{tabular}{lrrrr}
\toprule
\textbf{Approach} & \textbf{Acc.} & \shortstack{\textbf{Acc.}\\err=fail} & \shortstack{Inv.\\Cat.} & \shortstack{Ctx.\\Ovfl} \\
\midrule
Simple Sonnet (memo) & 31.9\% {\tiny[27,38]} & $=$ & -- & -- \\
Simple Sonnet (no-CoT) & 59.2\% {\tiny[54,64]} & 57.0\% {\tiny[52,62]} & -- & 5 (4\%) \\
Simple Sonnet & 59.7\% {\tiny[55,65]} & 57.0\% {\tiny[52,62]} & -- & 6 (4\%) \\
Simple Llama & 65.4\% {\tiny[62,71]} & 61.5\% {\tiny[56,66]} & 1 (1\%) & 7 (5\%) \\
Simple Mistral & 58.3\% {\tiny[54,64]} & 57.0\% {\tiny[52,62]} & -- & 3 (2\%) \\
Agent Sonnet (native) & 59.3\% {\tiny[54,64]} & $=$ & -- & -- \\
Agent Sonnet (stopseq) & 63.7\% {\tiny[59,68]} & $=$ & -- & -- \\
Agent Mistral & 58.5\% {\tiny[54,63]} & $=$ & -- & -- \\
\midrule
Random prediction\textsuperscript{$\dagger$} & 33.3\% & 33.3\% & -- & -- \\
Majority prediction\textsuperscript{$\ddagger$} & 36.3\% & 36.3\% & -- & -- \\
\end{tabular}
\end{table}

\subsubsection{H\"artel}
\tab~\ref{tab:accuracy-hartel} shows that our previous work on 
security defects is another harder task for all approaches.
We see accuracy in a range between 58\% and 66\% without counting errors as wrong classifications.
If context overflow errors are counted as wrong classifications,
the accuracy of all simple approaches drops below the agents. 
What is surprising is that Simple Llama comes out best when excluding errors,
almost coincidentally: it is the only approach that predicts `unclear' with a
non-negligible rate, which happens to be a good strategy for this task because
roughly a third of the samples are indeed labeled as unclear.

Our sanity check using Sonnet (memorization) shows an accuracy of 31.9\%,
which is below the accuracy of a random prediction (33.3\%) or a majority prediction (36.3\%).
We conclude that it is unlikely that the correct labels are in the training data.

Low performance might be caused by the original labels in~\cite{DBLP:conf/ease/Hartel25} being produced
without the pull request metadata. The resulting difference in available context, where our experiments have more
context, lets approaches avoid `unclear' and pick a category. We examine this in more detail in the
disagreement analysis in \sec~\ref{sec:disagreement-analysis}.

The comparison of agents against their simple counterparts in \fig~\ref{fig:accuracy-diff}
shows that the probability that any agent performs meaningfully worse
is low (\pWorseSonnetStopseqHartel{}, \pWorseSonnetNativeHartel{}, \pWorseMistralHartel{}
for stopseq, native, and Mistral).
Agent Sonnet (stopseq) even has a \pBetterSonnetStopseqHartel{} probability of being
meaningfully better than Simple Sonnet, which we attribute to
context overflow errors pulling Simple Sonnet accuracy down.

\begin{table}
    \centering
    \caption{Levin (maintenance activity, columns as in \tab~\ref{tab:accuracy-herbold})}
    \label{tab:accuracy-levin}
    \negvspacetab
    \footnotesize
    \setlength{\tabcolsep}{2pt}
\begin{tabular}{lrrrr}
\toprule
\textbf{Approach} & \textbf{Acc.} & \shortstack{\textbf{Acc.}\\err=fail} & \shortstack{Inv.\\Cat.} & \shortstack{Ctx.\\Ovfl} \\
\midrule
Simple Sonnet (memo) & 45.2\% {\tiny[39,52]} & 44.2\% {\tiny[38,51]} & 3 (2\%) & -- \\
Simple Sonnet (no-CoT) & 87.6\% {\tiny[82,91]} & $=$ & -- & -- \\
Simple Sonnet & 91.5\% {\tiny[87,94]} & $=$ & -- & -- \\
Simple Llama & 93.8\% {\tiny[89,96]} & $=$ & -- & -- \\
Simple Mistral & 89.9\% {\tiny[85,93]} & $=$ & -- & -- \\
Agent Sonnet (native) & 89.9\% {\tiny[85,93]} & $=$ & -- & -- \\
Agent Sonnet (stopseq) & 89.1\% {\tiny[84,92]} & $=$ & -- & -- \\
Agent Mistral & 89.1\% {\tiny[84,92]} & $=$ & -- & -- \\
\midrule
Random prediction\textsuperscript{$\dagger$} & 25.0\% & 25.0\% & -- & -- \\
Majority prediction\textsuperscript{$\ddagger$} & 50.4\% & 50.4\% & -- & -- \\
\end{tabular}
\end{table}

\subsubsection{Levin}
\tab~\ref{tab:accuracy-levin} shows numbers for maintenance activity classification. 
The original authors report 76\% accuracy for a model trained on commit messages. Our approaches are not
trained but reach accuracies between 87.6\% and 93.8\%, likely because LLMs encode
enough maintenance-activity knowledge to classify from the commit context directly.
Our sanity check with Sonnet (memorization) shows 45.2\% accuracy, below the
majority prediction with 50.4\% accuracy but above a random prediction
with 25.0\% accuracy. This means the class distribution may be somewhat reflected
in the training data, but the particular commit misses.

Comparing agents against their simple counterparts in \fig~\ref{fig:accuracy-diff},
all three pairs are practically equivalent
with \pEqualSonnetStopseqLevin{}, \pEqualSonnetNativeLevin{}, and \pEqualMistralLevin{}
probability of being within $\pm$5 accuracy points.
As with Herbold, no approach runs into errors on this task, so there is no robustness
advantage for agents to leverage.

\begin{table}
    \centering
    \caption{Munaiah (repository, columns as in \tab~\ref{tab:accuracy-herbold})}
    \label{tab:accuracy-munaiah}
    \negvspacetab
    \footnotesize
    \setlength{\tabcolsep}{2pt}
\begin{tabular}{lrrrr}
\toprule
\textbf{Approach} & \textbf{Acc.} & \shortstack{\textbf{Acc.}\\err=fail} & \shortstack{Inv.\\Cat.} & \shortstack{Ctx.\\Ovfl} \\
\midrule
Simple Sonnet (memo) & 82.0\% {\tiny[78,85]} & $=$ & -- & -- \\
Simple Sonnet (no-CoT) & 73.1\% {\tiny[68,77]} & 72.7\% {\tiny[68,77]} & -- & 1 (1\%) \\
Simple Sonnet & 83.6\% {\tiny[79,87]} & 83.1\% {\tiny[79,86]} & -- & 1 (1\%) \\
Simple Llama & 65.9\% {\tiny[60,70]} & 65.1\% {\tiny[60,70]} & -- & 2 (1\%) \\
Simple Mistral & 70.9\% {\tiny[66,75]} & $=$ & -- & -- \\
Agent Sonnet (native) & 87.8\% {\tiny[83,91]} & $=$ & -- & -- \\
Agent Sonnet (stopseq) & 84.3\% {\tiny[80,88]} & $=$ & -- & -- \\
Agent Mistral & 78.5\% {\tiny[74,82]} & $=$ & -- & -- \\
\midrule
Random prediction\textsuperscript{$\dagger$} & 33.3\% & 33.3\% & -- & -- \\
Majority prediction\textsuperscript{$\ddagger$} & 52.9\% & 52.9\% & -- & -- \\
\end{tabular}
\end{table}

\subsubsection{Munaiah}
\tab~\ref{tab:accuracy-munaiah} shows that repository classification for 
engineered software has the widest accuracy spread
across approaches, ranging from 65.1\% to 87.8\%. Our sanity check with Sonnet
(memorization) shows a striking 82.0\% accuracy, far above the majority
prediction with 52.9\% accuracy and the random prediction with 33.3\% accuracy.
These numbers suggest a partial training-data overlap for this task.
We cannot rule out that the model did see the labels, but we assume it
is more likely that repositories to be classified are present in the training data.
The trained knowledge on the repository might be sufficient to derive the label
just knowing the GitHub repository name, which is the only input to memorization.
Most surprising is that memorization beats Simple Sonnet (no-CoT),
Simple Llama, Simple Mistral, and Agent Mistral.

Comparing agents against their simple counterparts in \fig~\ref{fig:accuracy-diff},
Agent Mistral shows the clearest lean toward being meaningfully better
(\pBetterMistralMunaiah{} probability), while Agent Sonnet (native) has a
weaker \pBetterSonnetNativeMunaiah{} probability of being meaningfully better.
Only the stopseq pair is conclusively within the region of practical equivalence
(\pEqualSonnetStopseqMunaiah{} probability of being within $\pm$5 accuracy points).
This is the task with the strongest signal for an agent advantage,
primarily driven by Agent Mistral.

\subsection{Uncertainty Handling}

\begin{table}
    \centering
    \caption{Confusion matrix for `unclear' handling.}
    \label{tab:unclear-confusion}
    \negvspacetab
    \small
\begin{tabular}{@{}l rr rr@{}}
\toprule
 & \multicolumn{4}{c}{\textbf{Ground Truth}} \\
\cmidrule(l){2-5}
 & \multicolumn{2}{c}{Clear} & \multicolumn{2}{c}{Unclear} \\
\cmidrule(lr){2-3} \cmidrule(l){4-5}
\textbf{Approach} (concludes $\rightarrow$) & Clear & Unclear & Clear & Unclear \\
\midrule
Simple Sonnet (memo) & 389 & 0 & 44 & 0 \\
Simple Sonnet (no-CoT) & 549 & 9 & 83 & 1 \\
Simple Sonnet & 551 & 6 & 84 & 0 \\
Simple Llama & 512 & 42 & 67 & 17 \\
Simple Mistral & 550 & 10 & 83 & 2 \\
Agent Sonnet (native) & 562 & 0 & 86 & 0 \\
Agent Sonnet (stopseq) & 562 & 0 & 85 & 1 \\
Agent Mistral & 560 & 2 & 85 & 1 \\
\bottomrule
\end{tabular}
\end{table}

We show a detailed breakdown of which model predicts unclear regarding
whether the ground truth is clear or unclear in Tab.~\ref{tab:unclear-confusion}. The accuracy of correctly
matching unclear labels of the baseline is vanishingly low.
In total, models avoid classifying as unclear. 
Simple Llama uses \simpleLlmLlamaUnclearRate{} unclear labels,
simple Mistral \simpleLlmMistralUnclearRate{},
and simple Sonnet \simpleLlmSonnetUnclearRate{}.
We see that agents avoid labeling as unclear.
Our assumption is that broader access gives confidence to pick a category
and under-express uncertainty.
Overall, these numbers conform to the understanding that LLMs still struggle with uncertainty~\cite{DBLP:journals/corr/abs-2509-04664,DBLP:journals/tmlr/LinHE22}.

\subsection{Disagreement Analysis}
\label{sec:disagreement-analysis}

We have manually diagnosed \disagreementCasesLabeled{} of \disagreementCases{} disagreement cases
where at least two approaches (excluding memorization and no-CoT) disagree with the ground truth. 
\tab~\ref{tab:disagreement} summarizes by task and diagnosis. 
We discuss our findings in the following text and make the data available online to 
provide a starting point for refining the
original labeling guidelines we borrowed from the studies.

\begin{table}
    \centering
    \caption{Manual diagnosis of disagreements between approach 
    and the ground truth with number of labeled cases, disagreeing cases, and total cases per task.}
    \label{tab:disagreement}
    \negvspacetab
    \small
\begin{tabular}{@{}l rrrr rrr@{}}
\toprule
 & \multicolumn{4}{c}{\textbf{Diagnosis (labeled)}} & \multicolumn{3}{c}{\textbf{Cases}} \\
\cmidrule(lr){2-5} \cmidrule(l){6-8}
\textbf{Task} & Update & Keep & Spec. & Unres. & Lab. & Disag. & Total \\
\midrule
Herbold & 7 & 6 & 18 & 3 & 34 & 128 & 212 \\
Härtel & 20 & 1 & 6 & 1 & 28 & 59 & 135 \\
Munaiah & 4 & 9 & 9 & 1 & 23 & 52 & 172 \\
Levin & 6 & 1 & 7 & 1 & 15 & 15 & 129 \\
\midrule
\textbf{Total} & 37 & 17 & 40 & 6 & 100 & 254 & 648 \\
\bottomrule
\end{tabular}
\end{table}

\subsubsection{Herbold}

Disagreements predominantly trace to overlapping categories in the task specification.
A frequent ambiguity is between whitespace and refactoring, where formatting
adjustments can plausibly fall into either category. Similarly, when a commit
refactors documentation or test code, the correct label depends on whether the
\textit{type of change} (refactoring) or the \textit{type of artifact} (documentation, test)
takes precedence --- the guidelines do not resolve this.
A related pattern is test code that verifies the bugfix: is it part of the fix or a test?
Some \textit{unclear} ground-truth labels could be updated with additional context.

\subsubsection{H\"artel}
The majority of our analysis reflects that
our previous labeling was done without
pull-request metadata, such as the surrounding comments.
Additional context allows updating several \textit{unclear} labels. We further diagnosed specification issues 
with the phrase: `discusses a potential security defect'. This is unambiguous 
only if comments explicitly flag a security defect or are completely unrelated.
Comments that touch on security mechanisms without pointing to a defect 
split classifications. This is a specification issue, not an approach failure.

\subsubsection{Levin}

Levin shows low disagreements because of high accuracy.
Maintenance activity categories appear to be well-defined. 
If approaches disagree, three patterns emerge.
First, agentic approaches can recover context that the commit message misses.
In one of the cases, a commit message reads ``remove spam \ldots{} and fix a list.size == 0,'' yet the actual diff 
only replaces \texttt{list.size()==0} with \texttt{list.isEmpty()} and logging. This is a perfective cleanup and different from
ground truth classifying as a corrective fix. 
Second, approaches fall back on a repository's own classification taxonomy,
such as those present in the change logs.
This overrides a potentially better classification based on the actual change by the commit.
Third, test-only commits split approaches because the guidelines do not draw
a clear line between testing existing and new features.

\subsubsection{Munaiah}

Deciding between engineered projects and non-engineered projects leaves 
a wide gray zone. Missing thresholds on the amount of necessary documentation, single-event applications,
deprecated projects, and projects that have an architecture on the surface cause disagreement. Additional ambiguities arise for 
non-English naming conventions.

\section{Threats and Limitations}
\label{sec:threats_limitations}

\newcommand{\negvspacethreats}[0]{\vspace{-0.1cm}}

\paragraph{Data Leakage}

Leaked labels to LLM training might threaten our measurements.
We test for this by a memorization baseline, where three of four
tasks show negligible effects. Moreover, our comparison between agentic
and simple approaches is backed by the same LLMs, which factors out
such effects.

\negvspacethreats

\paragraph{Ground Truth Quality}

Biases and errors in the original labels can threaten
our accuracy measurements. We therefore manually diagnosed
\disagreementCasesLabeled{} disagreement cases, making
quality issues explicit.

\negvspacethreats

\paragraph{Setup Bias}

Specific prompt choices can influence behavior and threaten generalizability.
We standardize prompts across approaches and tasks, mirroring the original labeling guidelines. 

\negvspacethreats

\paragraph{Limitations of Sampling and Filtering}

Our findings are limited by the filters we apply to repository accessibility and size limits. This skews 
the sample toward projects that are smaller, active,
or still public and away from larger or archived repositories.

\negvspacethreats

\paragraph{Limited Generalizability to Newer Models}

We rely on models not publicly available, which implies that a part of our
experimental pipeline is out of our control. Different and newer releases may provide
different results in the future. Our findings are limited to the model versions we report on.
To compensate for this, we make the full trajectories available for a partial reproduction.

\negvspacethreats

\paragraph{Limited Feature Comparisons}

We limit our comparison to a range of basic agentic features against a simple LLM baseline across four tasks.
More advanced strategies, such as summarization strategies, agentic planning, truncation policies,
higher step limits, and internet access
for the agents, can clearly change our results.

\negvspacethreats

\paragraph{Limited Baseline Comparisons}
We limit approaches to not require task-specific engineering, except for the simple LLM baselines.
This disqualifies RAG approaches that rely on manually engineered retrieval strategies or any learning-based approaches. 

\section{Related Work}
\label{sec:related}

\smallskip

\subsection{Classification and Labeling in SE}

Authors of \cite{DBLP:journals/tse/YuTWM21} propose an active learning framework for vulnerability classification and labeling
called HARMLESS. It is a good example of pre-LLM vulnerability classification using a support vector machine.
Pre-LLM examples of explicit engineering include rule-based repository mining~\cite{DBLP:conf/ecmdafa/HartelHL18} and feature-based API clustering~\cite{DBLP:conf/iwpc/HartelAL18}.
Authors of~\cite{DBLP:journals/jss/AlfadelNCAS23} focus on classifying reviews of npm packages. 
Agentic classification techniques could potentially be applied here too, 
for a more profound analysis of contextual factors within the packages.
In~\cite{DBLP:conf/esem/YuFLTS23}, code reviews from the OpenStack and Qt communities are examined.
One conclusion is the benefit of combining manual review with automated techniques. 
In our task on security reviews, we take a step in this direction by examining how the review process
can be supported by LLM agents.
In~\cite{DBLP:journals/tosem/RahmanSBP23}, misconfigurations are examined.
Misconfigurations could be another interesting target for agentic classification.
Smells in infrastructure-as-code are examined in~\cite{DBLP:conf/icse/RahmanPW19}.
Plate et al.~\cite{DBLP:conf/icsm/PlatePS15} discuss that 
current decision-making based on natural-language vulnerability descriptions and expert knowledge 
is difficult, time-consuming, and error-prone. 

The approach we examine through contextualization by an agentic framework
is capable of working with arbitrary context, including natural-language descriptions, and all sorts of output that
git and Bash commands can provide. The agentic trajectories can also support semi-automated decision-making.

\subsection{LLMs and Agents in SE}

Commit messages that lack important information, potentially resolved by LLMs, are examined in~\cite{DBLP:conf/kbse/EliseevaSBGDB23}.
Assumptions on the context are related to ours in that knowing the repository history could 
significantly improve the quality of commit messages.
REPOAUDIT~\cite{DBLP:journals/corr/abs-2501-18160} combines static analysis with an LLM
to direct checks to suspicious code locations.
The agents we examine have flexible access to the commit history, using \code{git log}, \code{git diff}, and other commands,
allowing access beyond what is described before.
Recent work on the usage of LLMs for code review generation is presented in~\cite{DBLP:journals/corr/abs-2505-20206}.
Such work shows the current standard: code is passed as context to the LLM. Agentic techniques
that allow the LLM to flexibly retrieve code on their own are not yet considered.
In~\cite{DBLP:journals/tosem/YangCGLHLX25}, an empirical study on retrieval-augmented generation (RAG) for code generation is presented.
This study makes clear that RAG assumes `similar code snippets collected by a retrieval process' are the context.
We examine retrieval by agents that do not need a manually engineered \textit{similarity function} but 
a catalog of standard Bash commands to access a repository.
In~\cite{DBLP:conf/ease/AbeduAS24,DBLP:conf/msr/AbeduMKS25}, the authors use RAG and knowledge graphs to 
answer repository-related questions
by adding retrieved files to the prompt; they report that poor retrieval was the main cause of failures.
We see that issues with the similarity definition are common, and we bypass them by using standard Bash commands
to access the repository.

The first agentic framework for executing triage in software engineering is presented in~\cite{Triangle25}.
Cloud root-cause analysis with tool-augmented LLMs is explored in~\cite{DBLP:conf/cikm/WangLZZWYF0W24} 
where agents access monitoring data and logs autonomously to identify the root cause of cloud incidents.
Such approaches are close to what we examine but have a different focus.
SHERPA is described in~\cite{DBLP:journals/corr/abs-2509-00272} building on the idea of abstract state machines that sit on
top of the LLM invocations. Such ideas stem from seeing reasoning as a sequence of steps, a core for running an agent.

Moreover, recent MSR work studies agents themselves as subjects, analyzing traces they leave in repositories
like \code{AGENTS.md} or \code{CLAUDE.md}~\cite{MSR_Robbes26,MSR_Mohsenimofidi25,MSR_TasnimDasRoy26},
or differences in agent-generated code, such as in tests~\cite{MSR_HoraRobbes26}.
We take the complementary perspective: using agents to mine artifacts in software repositories.
\section{Conclusion}
\label{sec:conclusion}

We evaluated agentic repository mining, where LLMs classify repository artifacts by dynamically
exploring repository context through standard bash commands. Across four MSR tasks and eight approach
configurations, agents achieve competitive accuracy despite retrieving their own context,
while simple LLMs receive pre-engineered input. The primary advantage is robustness:
agents avoid context-window overflows and scale independently of artifact size.
A manual diagnosis of \disagreementCasesLabeled{} cases where approaches disagree with the
ground truth reveals specification ambiguities and labels produced under limited context,
suggesting that accuracy against such ground truth may underestimate approaches
with broader context access.

\begin{acks}
This work is funded by EU grant No. 101120393 (Sec4AI4Sec).
Anthropic's Claude Code (Opus 4.6) assisted coding and writing.
\end{acks}

\bibliographystyle{ACM-Reference-Format}
\bibliography{paper}

\end{document}